\documentclass[jap,twocolumn,showpacs,amsmath,letter]{revtex4}
\usepackage{graphicx}

\newcommand{\Journal}[4]{#1 \textbf{#2}, #3 (#4)}

\begin{document}

\title{Manipulating Current-Induced Magnetization Switching}

\author{S. Urazhdin}
\affiliation{Department of Physics and Astronomy, Johns Hopkins
University, Baltimore, MD, 21218}
\author{H. Kurt}
\author{M. AlHajDarwish}
\author{Norman O. Birge}
\author{W. P. Pratt Jr.}
\author{J. Bass}
\affiliation{Department of Physics and Astronomy, Center for
Fundamental Materials Research and Center for Sensor Materials,
Michigan State University, East Lansing, MI 48824}


\begin{abstract}
We summarize our recent findings on how current-driven magnetization switching and magnetoresistance in nanofabricated magnetic multilayers are affected by varying the spin-scattering properties of the non-magnetic spacers, the relative orientations of the magnetic layers, and spin-dependent scattering properties of the interfaces and the bulk of the magnetic layers. We show how our data are explained in terms of current-dependent effective magnetic temperature.
\end{abstract}
\maketitle

\section{\label{intro}Introduction}

Since the discovery of current-induced magnetization switching (CIMS) in magnetic nanopillars~\cite{tsoiprl,cornellscience}, researchers have investigated the effects of temperature~\cite{cornelltemp,myprl}, external magnetic field $H$~\cite{myprl,grollier,cornellquant}, layers' thicknesses~\cite{cornellquant}, interlayer interactions~\cite{myafcoupl,mycoupl}, and magnetic moment density~\cite{pufall}. The interest for such studies is twofold. For technological applications, it is necessary to minimize the switching currents for CIMS use in memory devices, or, on the contrary, minimize the current-induced noise in field sensors. From the fundamental point of view, CIMS allows one to access the strongly
out of equilibrium magnetic dynamics, and study the interaction of magnetization with conduction electrons and phonons. In Section~\ref{experiment}, we summarize our recent studies of how various modifications of the basic ferromagnetic-nonmagnetic-ferromagnetic (F$_1$/N/F$_2$) trilayer structure used for CIMS affect the switching currents $I_s$ and the magnetoresistance $\Delta R=R_{AP}-R_P$~\cite{iswvsmr,mustafa,myjmminpress}. Here, $R_{(AP)P}$ is the resistance of the state with (anti)parallel magnetizations of F$_1$ and F$_2$. First, we enhance $\Delta R$ in Py/Cu/Py (Py=Permalloy=Ni$_{84}$Fe$_{16}$) trilayer nanopillars by inserting $1$~nm of a strong spin-scatterer, Fe$_{50}$Mn$_{50}$~\cite{sdlength} between the trilayer and the top electrode. Second, we insert a $t_{CuPt}$ thick Cu$_{94}$Pt$_6$ layer between the Py layers. The short spin-diffusion length in Cu$_{94}$Pt$_6$ decreases $\Delta R$~\cite{sdlength}. Third, we study the variation of $\Delta R$ and $I_s$ with the angle between the magnetizations of the two ferromagnetic layers in Py/Cu/Py nanopillars. Fourth, we dope Ni or Fe with Cr for F$_1$ and/or F$_2$ to induce predominantly majority
electron scattering in the bulk of ferromagnets, or use Cr spacer with Fe to give predominantly majority electron 
scattering at F/N interface~\cite{cr}. Finally, in Section~\ref{model} we demonstrate that all our results are consistent with the effective magnetic temperature model (EMT) of Ref.~\cite{mytheory}.

\section{\label{experiment} Experiment}

Our samples were made with a multistep process described elsewhere~\cite{myafcoupl}. Below, all thicknesses are in nanometers. The basic samples had structure Cu/F$_1$/N/F$_2$/Au. In addition 
to the modifications specified for each sample below, in some samples a thin Cu spacer was inserted between
F$_2$ and the top Au. The Cu/F$_1$ layers were thick extended bottom leads. N and  F$_2$ were
patterned into an elongated shape with typical dimensions $\approx 130\times 70$~nm (nanopillar). Au was a thick top lead. Leaving F$_1$ extended minimized the effect of dipolar coupling on the current-driven switching~\cite{myafcoupl}, and N was sufficiently thick to minimize exchange coupling between F$_1$ and F$_2$. We measured $dV/dI$ at room temperature (295~K) with four-probes and lock-in detection, adding an ac current of amplitude 20--40~$\mu$A at 8~kHz to the dc current $I$. Positive current flows from the extended to the patterned Py layer.
The field $H$ is in the film plane and (except for the angular dependence studies) along the nanopillar easy axis.

\subsection{Enhancing CIMS}

\begin{figure}
\includegraphics[scale=0.8]{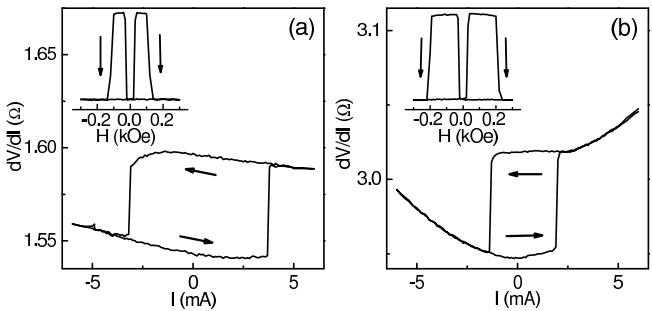} \caption{\label{femn} (a) $dV/dI$ {\it vs.} $I$ for a sample of type 1 (as defined in the text) at $H=0$. Inset: $dV/dI$ {\it vs.} $H$ at $I=0$. (b) Same as (a), for a sample of type 2. (From Urazhdin {\it et al.}~\cite{iswvsmr})}
\end{figure}

Fig.~\ref{femn}(a) shows typical CIMS results for a sandwich 
F$_1$=Py(30)/N=Cu(15)/F$_2$=Py(6)/Cu(2) (sample of type 1). In samples of type 2, the top Cu(2) layer was replaced with a Cu(2)/Fe$_{50}$Mn$_{50}$(1)/Cu(2) sandwich (Fig.~\ref{femn}(b)).
The typical switching fields, $H_s$ (insets Fig.~\ref{femn}) were similar for both sample types. For 14 samples of type 1, $\Delta R\equiv R_{AP}-R_{P}=0.060\pm 0.002\Omega$. $I_s^{AP\to P}=-2.45\pm 0.2$~mA, and $I_s^{P\to AP}=3.8\pm 0.2$~mA are the switching currents from AP to P, and from P to AP state, respectively. For 12 samples of type 2, $\Delta  R=0.085\pm 0.012\Omega$,
$I_s^{AP\to P}=-1.5\pm 0.2$~mA, and $I_s^{P\to AP}=1.85\pm 0.2$~mA. For uncertainties, we give twice the standard deviations of the mean. Thus, insertion of FeMn outside the Py/Cu/Py trilayer
reduces the average $I_s^{AP\to P}$ by a factor of $\approx 1.6$, and $I_s^{P\to AP}$ by $2.1$. The increase of $\Delta R$ results from the enhancement of spin-accumulation in the trilayer, due
to the spin-memory loss in FeMn(1)~\cite{sdlength}. The associated decrease of $I_s$ indicates
the importance of spin-accumulation for CIMS.

\subsection{Suppressing CIMS}

\begin{figure}
\includegraphics[scale=0.8]{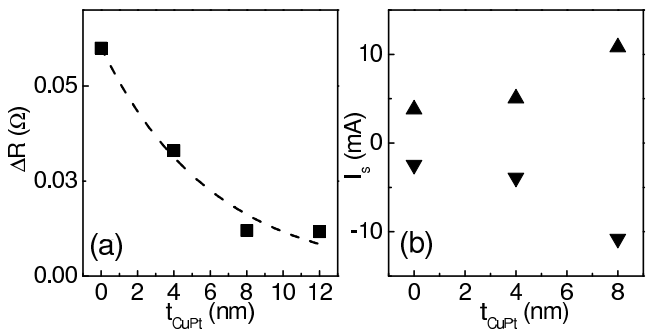}
\caption{\label{cuptall} (a) Variation of $\Delta$R with $t_{CuPt}$. Dashed line is a fit with $\Delta R=\Delta R_0exp[-t_{CuPt}/l_{sf}^{CuPt}]$, $\Delta R_0=0.060\pm 0.004\Omega$, $l_{sf}^{CuPt}=6.1\pm 0.8$~nm. (b) $I_s^{P\to AP}$ (upward triangles) and $I_s^{AP\to P}$ (downward triangles) {\it vs.} $t_{CuPt}$. (From Urazhdin {\it et al.}~\cite{iswvsmr})}
\end{figure}

Sample types 3-5 had the structure of sample type 1, except N=Cu(15)
was replaced with Cu(13.5-d)/Cu$_{94}$Pt$_6$(d)/Cu(1.5), with d= 4, 8, and 12, respectively.
Fig.~\ref{cuptall}  shows data for sample types 1, 3, 4, 5. Fig.~\ref{cuptall}(a) ($\Delta$R($t_{CuPt})$) gives a spin-diffusion length of $6.1\pm 0.8$~nm in Cu$_{94}$Pt$_6$ at 295~K, shorter than $\approx10$~nm at 4.2~K~\cite{sdlength}. Fig.~\ref{cuptall}(b) shows that both $I_s^{P\to AP}$ and $|I_s^{AP\to P}|$ increase with increasing $t_{CuPt}$. Interestingly, the ratio $Q=I_s^{P\to AP}/|I_s^{AP\to P}|$ decreases from $\approx 1.5$ at $t_{CuPt}=0$ to $\approx 1.0$ for $t_{CuPt}=8$. This decrease with increased  spin-flipping within the N-layer is opposite
to that reported in~\cite{cornellquant} for a similar measurement with varied thickness of N=Cu,
and is inconsistent with the explanation proposed there.

\begin{figure}
\includegraphics[scale=0.8]{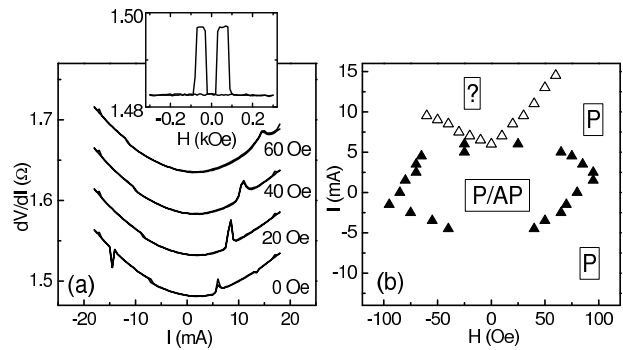}
\caption{\label{cupt12nm} Data for a sample of type 5: (a) $dV/dI$ {\it vs.} $I$. Inset: $dV/dI$ {\it vs.} $H$. (b) Stability diagram, obtained from H-scans at various fixed I (solid symbols), and I-scans at fixed H (open symbols). P is the region where only the P state is stable, P/AP is a bistable region, and the region labeled [?] is the inhomogeneous magnetization state.
From Urazhdin {\it et al.}~\cite{myjmminpress}}
\end{figure}

The inset of Fig.~\ref{cupt12nm}(a) illustrates that all 8 samples of type 5 showed hysteretic field-driven switching, similar to the other sample types. However, none exhibited reproducible hysteretic current-switching. As Fig.~\ref{cupt12nm}(a) shows for $H=0$ to $60$~Oe, the current dependencies exhibit reversible switching peaks at small $H$, rapidly moving to higher $I$ at larger $H$. The resistance change associated with such peaks indicates switching between the P state and an inhomogeneous magnetization state, likely affected by the current-induced Oersted field. In Fig.~\ref{cupt12nm}(a), P to inhomogeneous state switching without a peak also occurs at $H=0$, $I=-7$~mA, with a partial reverse transition at $I=-14.5$~mA (downward peak). In some of the other samples of type 5, a switching peak appears only at $I<0$, or in both directions of $I$. Averaging among 8 samples, at $H=0$ the peaks appeared at $|I|\approx 10$~mA. Fig.~\ref{cupt12nm}(b) shows
why CIMS is absent in these samples: When $I$ is reduced from a large positive value at small $H$,
the magnetization M$_2$ of F$_2$ switches reversibly from the inhomogeneous to the P state
(as $I$ crosses the line marked by open symbols), then goes through the P/AP and P regions
without changing. 
The AP state is  achieved only by flipping the magnetization M$_1$ of $F_1$ with $H$. The results of Fig.~\ref{cupt12nm} are important in the context of attempts to understand the current-driven behavior of a single magnetic layer~\cite{polianski,tingyong,kent}. By placing a thick CuPt layer in the spacer between two Py layers, we approached a single-layer regime, while retaining a small $\Delta R$ necessary to determine
the orientation of M$_2$. An approximate symmetry of the effects of positive and negative currents (Fig.~\ref{cupt12nm}) shows that, indeed, the switching of M$_2$ is not significantly
affected by M$_1$.

\subsection{Noncollinear Switching}

\begin{figure}
\includegraphics[scale=0.8]{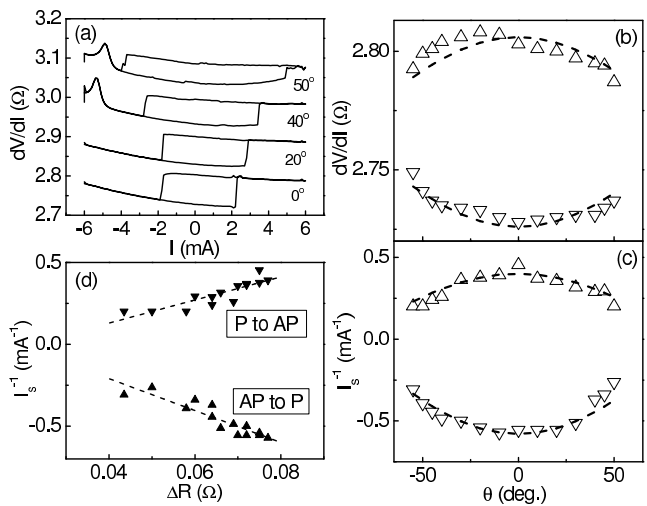}
\caption{\label{amr} (a) CIMS for various non-collinear orientations of M$_1$ and M$_2$, as labeled. Curves are offset for clarity. (b) Quasi-parallel (P) and quasi-antiparallel (AP) state resistances $R_P$ (upward triangles) and $R_{AP}$ (downward triangles) {\it vs.} $\theta$. Dashed curves: fits as explained in the text. (c) $1/I_s^{P\to AP}$ (upward triangles) and $1/I_s^{AP\to P}$ (downward triangles) {\it vs.} $\theta$. Dashed curves: fits as explained in the text. (d) $1/I_s$ vs. $\Delta R$ for noncollinear switching, dashed lines are best linear fits.} 
\end{figure}

Fig.~\ref{amr} shows the results for varied non-collinear orientations of magnetic layers in sample type 1.  These data confirm results reported in~\cite{mancoff}, but with a larger magnetoresistance $\Delta R/R$.
Before each measurement, a pulse of $H=60$~Oe at the desired in-plane angle $\theta$ was applied to rotate M$_1$ parallel to $H$, then the current-switching was measured at $H=10$~Oe, needed to fix M$_1$. Fig.~\ref{amr}(a) shows that for $\theta<60^\circ$ the switching occurred though a well
defined sharp step, at increasing $I_s$ and decreasing $\Delta R$
with increasing $\theta$. At larger $\theta$, the switching points became poorly defined. They are
not included in Fig.~\ref{amr}. Fig.~\ref{amr}(b) shows that the resistance of the quasi-parallel state increased,  and the quasi-antiparallel state decreased when $|\theta|$ increased. Dashed curves
are fits assuming $R_{P,AP}=R_0\mp A\cos\theta$, $R_0=2.77\Omega$, $A=0.04\Omega$,
as justified in Section~\ref{model}. $\Delta R\approx 2A=0.08\ \Omega$ for the collinear magnetoresistance
of this sample. Dashed curves in Fig.~\ref{amr}(c) for the inverse switching currents are fits 
with $1/I_s^{P\to AP,AP\to P}=K_{P,AP}\cos\theta$, $K_P=0.40$~mA$^{-1}$, $K_{AP}=-0.58$~mA$^{-1}$. Here $K_P\approx 1/I_s^{P\to AP}$ and $K_{AP}\approx 1/I_s^{AP\to P}$ for the CIMS with collinear
magnetizations. Dashed lines in  Fig.~\ref{amr}(d) for the $1/I_s$ vs. $\Delta R$ are best linear fits.
Their intercepts with $1/I_s=0$ axis are at finite $\Delta R$.

\begin{figure}
\includegraphics[scale=0.5]{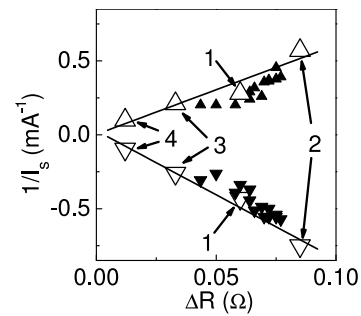}
\caption{\label{isvvsmrall} Dependence of $1/I^{P\to AP}_s$ (upward triangles) and $1/I^{AP\to P}_s$ (downward triangles) on $\Delta R$. Open symbols: sample types 1 through 4, as labeled. Solid symbols: variations with angle between the magnetizations in a sample of type 1. Solid lines: best linear fits of data, excluding the angular dependence.(From Urazhdin {\it et al.}~\cite{iswvsmr})}
\end{figure}

In Fig.~\ref{isvvsmrall}, we plot average $1/I_s$ {\it vs.} average $\Delta$R
using the data of Figs.~\ref{femn}-\ref{amr}. The uncertainties are close to the
symbol sizes in Fig.~\ref{isvvsmrall}. The overall agreement for three different types of measurements suggests a general inverse relationship between $I_s$ and $\Delta R$, independent of the particular way in which $\Delta R$ was varied. The switching is determined by the current density, so both $1/I_s$ and $\Delta R$ are inversely proportional to the nanopillar areas; their variation only leads to scaling along the approximately linear dependence in Fig.~\ref{isvvsmrall}(a). The solid lines show best linear
fits, excluding the angular dependence data. The ordinate intercepts of both fits are zero
within the uncertainty of the fits.
 
\subsection{Inverting Magnetoresistance and/or CIMS}

\begin{figure}
\includegraphics[scale=0.5]{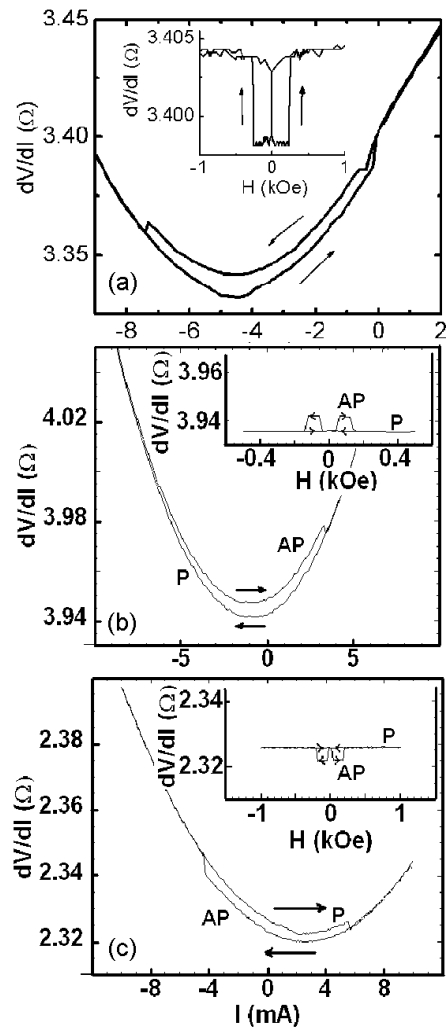}
\caption{\label{invall} (a-c) $dV/dI$ {\it vs.} $I$ for samples of types 6-8, correspondingly.
Insets: $dV/dI$ {\it vs.} $H$. (b), (c) are from AlHajDarwish {\it et al.}~\cite{mustafa}.}
\end{figure}

In sample types 1-5, the minority electrons (i.e. with moments anitparallel to the layers' magnetizations)
were always scattered more strongly both in the bulk of F$_1$ and F$_2$ and at their interfaces. 
Fig.~\ref{invall} demonstrates the effects  on $\Delta R$ and CIMS of inverting the scattering
anisotropies~\cite{mustafa}.

Fig.~\ref{invall}(a) is for F$_1$=Ni$_{97}$Cr$_3$(20)/N=Cu(20)/F$_2$=Py(10) (sample type 6). By doping Ni with Cr, we inverted the anisotropy of F$_1$, so that majority electrons are scattered in F$_1$ more strongly than the minority ones. Inset of Fig.~\ref{invall}(a) shows that $\Delta R$ is inverted (as compared to normal $\Delta R$ in samples of type 1), i.e. $R_P>R_{AP}$. CIMS in Fig.~\ref{invall}(a) looks qualitatively similar to Fig.~\ref{femn},
but inverted $\Delta R$ means that CIMS is actually also inverted, i.e. $I>0$ gives switching to the high resistance P state, and $I<0$ gives switching to the low resistance AP state. 

Fig.~\ref{invall}(b) is for F$_1$=Fe$_{95}$Cr$_5$(20)/N=Cr(6)/F$_2$=Fe$_{95}$Cr$_5$ (sample type 7). The bulk and interfaces
of ferromagnets with Cr all scatter the majority electrons more strongly. Inset of Fig.~\ref{invall}(b) shows that the sign of $\Delta R$ is the same as 
for samples of type 1, as is always the case for symmetric trilayers. CIMS in Fig.~\ref{invall}(b) is inverted.

Fig.~\ref{invall}(c) is for F$_1$=Py(20)/N=Cu(7)Cr(3)/F$_2$=Fe$_{95}$Cr$_5$ (sample type 8). In this case, F$_1$ and F$_1$/N have positive
anisotropy, but F$_2$ and F$_2$/N have negative anisotropy. Inset in Fig.~\ref{invall}(c) shows that $\Delta R$ is inverted. CIMS also looks inverted compared to samples of type 1,
but since $\Delta R$ is inverted, CIMS is actually the same, i.e. $I>0$ gives the low-resistance AP state, and $I<0$ gives the high-resistance P state.

In summary, samples 1,6-8 demonstrate all 4 possible combinations of signs of $\Delta R$ and CIMS: labeling these signs normal in sample type 1, 
samples of type 6 show inverted $\Delta R$ and inverted CIMS, samples of type 7 show normal $\Delta R$ and inverted CIMS, and samples of type 8 show inverted $\Delta R$ and normal CIMS.

\section{\label{model} Model and Analysis}

The generally accepted spin-transfer torque model (STT) for CIMS is usually combined with the macrospin current-driven magnetization dynamics~\cite{slonczewski,berger}. Finite-temperature analysis showed that the effect of STT on the macrospin dynamics is equivalent to renormalization of temperature T, giving~\cite{kochsun,zhang}
\begin{equation}\label{tprime}
T'(I)=\frac{T}{1+pI'\hbar/(2em\alpha H_{eff})},
\end{equation}
where $\alpha$ is the Gilbert damping parameter, $H_{eff}\approx H+2\pi M_2$, and $m$ is the magnetic moment of F$_2$.
We show elsewhere~\cite{mytheory,myunpublished} that $I'$ should be generally understood as the 
current through the nanopillar if electron scattering on F$_2$ were absent, and $p$ is that current's polarization.
$T'$ was treated in Refs.~\cite{kochsun,zhang} as a formal parameter for the current-dependence of  the switching rate
$\Omega(I)=\Omega_0exp[-U/kT'(I)]$, where $\Omega_0\approx 10^{-9}$~sec is the switching attempt rate, and $U$ is the switching
barrier determined by the anisotropy field of the nanopillar. Furthermore, it can be shown that $T'$ is the actual magnetic
temperature of the macrospin, defined by the probability distribution of its orientation~\cite{zhang,myunpublished}.

STT has been recently extended to inhomogeneous dynamical states~\cite{polianski} and incorporated in micromagnetic models~\cite{zhu,miltat,lee}. Micromagnetic simulations invariably show that CIMS occurs through inhomogeneous intermediate magnetic states, not adequately described by the macrospin approximation even in nanostructures. We introduced effective magnetic temperature (EMT)
to collectively describe such inhomogeneous dynamical states~\cite{myprl,mytheory}. Surprisingly,
an apparently different approximation used by EMT gives an expression similar to Eq.~\ref{tprime}.
We show elsewhere~\cite{myunpublished} that contrary to the claim of Ref.~\cite{mytheory}, this similarity is not accidental: incoherent excitation of finite wavelength modes simply gives larger $H_{eff}$ in Eq.~\ref{tprime},
reflecting higher average damping rates of finite-wavelength excitations. We shall see below that macrospin calculations underestimate the switching currents by an order of magnitude, or equivalently require
unrealistically large Gilbert damping $\alpha$, to compensate for the higher $H_{eff}$ associated with nonuniform magnetic dynamics.

Here, we use a simplified quasi-ballistic approach for electron scattering in the nanopillar, and Eq.~\ref{tprime} to explain our
data. Following Ref.~\cite{mytheory}, we approximate the nanopillar by a high-resistance constriction separating low-resistance reservoirs. Only $F_2$ is contained in the constriction, $F_1$ and various nonmagnetic inserts such as Cu$_{94}$Pt$_6$ or Fe$_{50}$Mn$_{50}$, if present, are part of the reservoirs. To model the polarizing effect of F$_1$,  we introduce different spin-up and spin-down potentials in the left reservoir containing $F_1$, $V^{\uparrow}$, $V^{\downarrow}$, which are determined by the magnetic orientation F$_1$ and the properties of spacer, e.g. spin-memory loss in Cu$_{94}$Pt$_6$. This approximation is equivalent to
introducing different numbers of channels for spin-up and spin-down electrons in Landauer-Buttiker
formalism, physically justified by the spin-dependent band structure of F$_1$. We neglect spin-accumulation in the right reservoir, i.e. take $V=0$ there. $I^{\uparrow}$, $I^{\downarrow}$ are spin-up and spin-down currents through the constriction. Positive currents are from F$_1$ to F$_2$. Spin states are defined with respect to the magnetization M$_2$ of F$_2$. We define $\Delta V$, $p$ by $\Delta V=V^{\uparrow}-V^{\downarrow}=2pV$. Here $V\approx (V^{\uparrow}+V^{\downarrow})/2$ is somewhat smaller than the voltage across the multilayer because of the current bunching in the leads. If F$_2$ is removed, $p$ becomes the current polarization in the constriction, positive for the P state, and negative for the AP state for Py-based nanopillars. For simplicity, we approximately describe the contribution of F$_1$ and its interfaces and other various contact and bulk contributions to the nanopillar resistance by a resistance $R_0$. Spin-dependent scattering at each of the interfaces of F$_2$ is described by conductances $G^{\uparrow},G^{\downarrow}$, so that $V^{\uparrow(\downarrow)}=I^{\uparrow(\downarrow)}(2R_0+2/G^{\uparrow(\downarrow)})$. We neglect multiple scattering, so the interface resistances add. For now we also assume no spin-flip scattering in $F_2$. In this approximation,
\begin{equation}\label{v}
V=\frac{I}{(\frac{1+p}{2R_0+\frac{2}{G^\downarrow}}+\frac{1-p}{2R_0+\frac{2}{G^\uparrow}})}
\end{equation}
and 
\begin{equation}\label{mr}
\Delta R=\frac{4p(R_0+1/G^\uparrow)(R_0+1/G^\downarrow)(1/G^{\downarrow}-1/G^\uparrow)}
{[2R_0+1/G^{\downarrow}+1/G^\uparrow]^2-p^2[1/G^{\downarrow}-1/G^\uparrow]^2},
\end{equation}
approximately proportional to $p$. 

Similarly to STT~\cite{slonczewski}, we assume that all the electrons scattered by F$_2$ contribute equally to spin-transfer, regardless whether they are reflected or transmitted
by F$_2$. Only the transmitted electrons contribute to the charge current, so 
we need to find the relation between the current and the number of electrons scattered (i.e. either reflected or transmitted) by F$_2$. In Eq.~\ref{tprime} for spin-transfer, $pI'$
is the polarized current through the constriction if F$_2$ were absent,
giving $pI'/e$ - the difference between the numbers of electrons with spin-up and spin-down, scattered (i.e. either reflected or transmitted) 
by F$_2$ per unit time. Plugging the voltage from Eq.~\ref{v} into $pI'=pV/R_0$, \begin{equation}\label{etaI}
pI'=\frac{2pI(R_0+1/G^\downarrow)(R_0+1/G^\uparrow)}{R_0[2R_0+1/G^\uparrow+1/G^\downarrow +p(1/G^\downarrow -1/G^\uparrow )]}
\end{equation}
We shall see below that the spin-dependent scattering by F$_2$ explains the asymmetry of the switching currents. 

In EMT, CIMS occurs through thermal activation over an effective magnetic barrier determined by the shape anisotropy of the nanopillar~\cite{mytheory}. For the average switching time $\tau=1$~sec, the switching temperature is $T_0(I)\approx U/(k_B\ln\Omega_0)$, where $\Omega_0\approx 10^9$~sec$^{-1}$ is the effective attempt rate~\cite{kochsun}, and $U\approx 0.9$~eV is the estimated activation
barrier for $6$~nm thick Py nanopillars at 295~K. Approximating $U(T)\approx\sqrt{1-T/T_c}$, where $T_c=800$~K for Py, we obtain $T_0\approx 430$~K. We calculate
the switching currents from Eq.~\ref{tprime} with $pI'$ from Eq.~\ref{etaI}
\begin{eqnarray}\label{isw}
\nonumber  1/I_s=\frac{\hbar}{em\alpha H_{eff}(1-T/T_0)}\times\\
\frac{ p(R_0+1/G^\downarrow)(R_0+1/G^\uparrow)}{
R_0[2R_0+1/G^\uparrow+1/G^\downarrow +p(1/G^\downarrow -1/G^\uparrow )]}
\end{eqnarray}
Eq.~\ref{isw} shows that $I_s$ is approximately inversely proportional to the current polarization created by $F_1$. The asymmetry of $I_s^{P\to AP},I_s^{AP\to P}$ is determined by the values of $R_0$, $G$, and $p$, and is a consequence of the spin-dependent scattering by the nanopillar. The fundamental quantity giving CIMS is not the current, but rather the 
spin-dependent voltage across the nanopillar.

We perform numerical analysis with the derived equations using the statistically significant data
for sample types 1-5, and give only qualitative explanations for sample types 6-8.
We start with the ratio $Q=I_s^{P\to AP}/|I_s^{AP\to P}|$, which lets us find the ballast resistance $R_0$. For our Py nanopillars with area $\approx 10^{-14}m^2$, $G^{\uparrow}=33\pm 3$~s, $G^{\downarrow}=6\pm 0.6$~s~\cite{pratt,nadgorny}. For Py/Cu/Py samples, we estimate $p_0=0.6$. For $R_0=0$, Eq.~\ref{isw} gives $Q=2.4$. For large $R_0$, $Q=1$. $R_0=0.1\ \Omega$ gives $Q=1.5$ in agreement with experiment. This value is comparable to the Sharvin resistance of the nanopillar. For samples with Cu$_{94}$Pt$_6$ inserts, $p\approx p_0exp[-t_{CuPt}/l_{sf}^{CuPt}]$ with $l_{sf}^{CuPt}=6.1\pm 0.8$~nm (Fig.~\ref{cuptall}). Eq.~\ref{isw} then gives a decrease of $Q$ from $1.5$ for samples of type 1 to $1.1$ in samples of type 4, in reasonable agreement with the data of Fig.~\ref{cuptall}.

Plugging realistic sample parameters into Eq.~\ref{mr} gives $\Delta R=0.07\Omega$ for samples of
type 1, in good agreement with the data. The deviations from the linear dependence on $p$ in
Eq.~\ref{mr} are negligible, thus also reproducing the dependence of $\Delta R$ on $t_{CuPt}$ for sample types 3-5.

The calculation of $I_s$ using Eq.~\ref{isw} requires estimates for the Gilbert damping parameter $\alpha$, which in thin films may significantly deviate from the bulk value, and for the parameter $H_{eff}$, which depends on the typical energies of the magnetic modes excited by the current.
For CIMS in nanopillars at small in-plane $H$, $H_{eff}$ for uniform precession becomes
$H_{eff}\approx 2\pi M+H$, where $M\approx 800$~Oe is the magnetization of Py.
For $H\ll 2\pi M$, $H_{eff}\approx 5$~kOe. The $I_s$ data of
Sec.~\ref{experiment} are reproduced with unrealistic $\alpha\approx 0.12$. Thin-film FMR measurements indicate that for our $6$~nm thick Py
the damping should not significantly exceed the bulk value $\alpha\approx 0.01$~\cite{urban}. Eq.~\ref{isw} then gives $H_{eff}\approx 5$~T, corresponding to typical energies $E_i\approx 0.7$~meV for magnons participating in the magnetization switching. Incidentally, this estimate is consistent with the conduction electron energies due to the voltage across the nanopillar. This is expected from energy conservation in the electron-magnon scattering, not considered in the STT-based models. 

The corrections to the linear dependencies on $p$ in Eqs.~\ref{mr},~\ref{isw} are negligible. 
Thus, we reproduce the inverse relationship $1/I_s\propto\Delta R\propto p$ for sample types 1,3,4,
where $p$ was varied by inserting Cu$_{94}$Pt$_6$. By respective rotations of the layers' magnetizations
in sample type 1 (Fig.~\ref{amr}), we varied $p\approx p_0\cos\theta$, where  $p_0$ is the polarization for the
collinear magnetic orientation. Eqs.~\ref{mr},~\ref{isw} then validate the functional forms we used for fitting those data.

We interpret the behaviors of Fig.~\ref{cupt12nm} with two alternative explanations: i) the dominant effect of the unpolarized current through the nanopillar is due to its Oersted field and/or Joule heating. We speculate that these effects reduce the value of $H_s$, independently of the current direction; ii) unpolarized current spontaneously generates magnetic excitations, at a rate that is independent of both the current direction and the orientation of the
nanopillar~\cite{mytheory}. More detailed experiments and modeling are necessary to determine
the relative importance of the alternatives i) and ii).

Finally, the behaviors of sample types 6-8 with inverted anisotropies are explained by Eqs.~\ref{mr},~\ref{isw}, keeping in mind that $p$ is the polarization of current
that would flow through the nanopillar if F$_2$ were absent. From Eq.~\ref{mr},
$\Delta R\propto p(1/G^\downarrow-1/G^\uparrow)$, positive when anisotropies of F$_1$ and F$_2$
are the same (sample types 1 and 7), and negative when they are opposite (sample types 6,8),
From Eq.~\ref{isw}, the switching current sign only depends on the anisotropy of F$_1$, i.e.
the sign of $p$: positive in sample types 1 and 8, giving normal CIMS, and negative for sample
types 6 and 7, giving inverted CIMS. Our model does not include the spin-accumulation effects
caused by the anisotropy of F$_2$~\cite{mustafa}. These effects are most pronounced in samples
of type 6 with a highly anisotropic F$_2$=Py and weakly anisotropic F$_1$=Ni$_{97}$Cr$_3$,
giving asymmetry of the currents for the opposite switching directions.

\section{Summary}

In this paper, we summarized the following important experimental
observations for the variation of the magnetization switching current $I_s$ with 
multilayer parameters:
i) $I_s$ is reduced by inserting a strongly spin-scattering layer between 
the magnetic trilayer and one of the leads, ii) $I_s$ is increased by inserting a
spin-scattering layer between the magnetic layers, iii) $I_s$ is smallest when the
magnetizations of the layers are collinear, iv) for a fixed switching layer, there is an approximately linear relation between the magnetoresistance and $1/I_s$, regardless of the method by which these parameters are varied, v) all four possible combinations of magnetoresistance and switching current signs are
produced by manipulating with the scattering anisotropies of the magnetic layers and their interfaces. In our samples, the sign of the switching current is determined by the overall anisotropy of the fixed magnetic layer, and independent of the anisotropy of the switching layer. Finally, we show that our data are consistent with the predictions of the effective magnetic temperature model.

This work was supported by the MSU CFMR, CSM, the MSU Keck
Microfabrication facility, the NSF through Grants DMR 02-02476,
98-09688, and NSF-EU 00-98803, and Seagate Technology.

\end{document}